\documentclass[12pt]{iopart}

\usepackage{iopams}
\expandafter\let\csname equation*\endcsname\relax
\expandafter\let\csname endequation*\endcsname\relax
\usepackage{amsmath, amsthm, amssymb, slashed, graphicx, float, subcaption, appendix, color, ulem}

\begin{document}

\title[Resonant Scattering of Gravitational Waves With Electromagnetic Waves]{Resonant Scattering of Gravitational Waves With Electromagnetic Waves}

\author{ Ruodi Yan${}^1$, Yun Kau Lau ${}^2$\footnote{Corresponding author, email lau@amss.ac.cn}.}

\address{${}^1$ Department of Physics, Beijing Normal University, No.19, Xinjiekouwai Street, Haidian District, Beijing, P.R.China,}
\address{$^2$  Institute of Applied Mathematics, Morningside Center of Mathematics, LSSC,  Academy of Mathematics and System Science, Chinese Academy of Sciences, 55, Zhongguancun Donglu, Beijing, 100190, China.}
\vspace{10pt}


\begin{abstract}
A certain class of exact solutions of  Einstein Maxwell spacetime in general relativity is discussed which demonstrates at the level of theory that, when certain parametric resonance condition is met, the interaction of electromagnetic field with a gravitational wave will display certain Liapounov instability and lead to exponential amplification of a gravitational wave train described by certain Newman-Penrose component of the Weyl curvature. In some way akin to a free electron laser in electromagnetic theory, by the conversion of electromagnetic energy into gravitational energy in a coherent way,  the feasibility of generating a pulsed laser like intense beam of gravitational wave is displayed. 
\end{abstract}

%

%
%
%
%

\section{Introduction}

In electromagnetic theory, an intense light beam, known to have a diverse number of
practical applications in science and engineering, usually is generated by a laser whose
operating principle depends on the quantum stimulated emission of photons by molecules
in a medium. A shortcoming of this kind of laser source is that the frequency of the light
beam emitted depends heavily of the energy levels of the kind of molecules employed and
as a result the tunable frequency range is severely limited. One way to overcome this
spectral limitation is to employ free electron laser whose operating principle is classical
in nature. The underlying physics hinges on the resonant interaction between the electromagnetic field generated by a highly
relativistic electron beam and that of an oscillating magnetic field. When the frequencies of the electromagnetic field of the electron beam and that generated by an
oscillating magnetic field come close to each other, resonant interaction occurs between the two electric fields and an intense beam of light is generated\cite{rabinovich2012oscillations}.

The aim of the present work is to show that, within the context of Einstein’s
general theory of relativity, when certain resonance condition is met, a spacetime describing the interaction between gravitational wave and electromagnetic field 
will display certain Liapounov instability in a sense to be described in what follows. The
pumping of electromagnetic energy into a gravitational field will also generate an
intense beam of gravitational waves. It is known that the ergosphere of a rotating black hole is able to amplify gravitational wave by means of an electromagnetic field \cite{starobinskii1973amplification}. Our work presents another way of amplification without necessarily the mediation of a black hole.

Though the underlying operational principles behind a free electron laser and that presented in this work are different, both share the common feature that certain resonance condition is required to generate an intense beam. On the basis of this loose analogy and, as we shall see later,  the coherent conversion between gravitational and electromagnetic energy, the beam of intense gravitational wave may be termed “gravilaser”. It should be admitted that the class of spacetimes considered is highly idealistic and quite remote from experimental realisation and practical applications in the near future, still it demonstrates that at the level of theory that such a mechanism of generating an intense beam of gravitational wave train in laboratory is allowed by general relativity. This is in contrast to the general belief that gravitational wave generated in laboratory is too weak to be considered (see for instance\cite{saulsonFundamentalsInterferometricGravitational2017}).

The paper is structured as follows. The class of spacetimes to be considered in the
present work will be described in section 2. Section 3 contains a theoretical discussion
on the parametric resonance (or Liapounov instability) phenomena originated from the
Einstein Maxwell field equations and the nature of spacetime singularity of the class of
spacetime concerned. Numerical simulation will then be presented in section 4 to illustrate in details
how an intense beam of pulse gravitational wave train is generated by the coherent
conversion of electromagnetic energy into gravitational energy and the waveform of the
intense beam. This will then be followed by some brief remarks in section 5 to conclude our work
in the final section.

\section{Background and notations. }
\label{sec:bg}

In this section we will give a brief description of the class of spacetime to be considered in what follows and on the basis of which we develop our work. 

Consider a spacetime whose metric is given as \cite{liangFamilyCylindricallySymmetric1995,kuangEinsteinMaxwellSpacetimeTwo1999}
\begin{eqnarray}\label{metric}
    ds^2 = 2e^{2\zeta} du dv - e^{2\eta} \cosh{\omega} (dx^2 + e^{2\chi} dy^2) - e^{2\eta + \chi} \sinh{\omega} dx dy.
\end{eqnarray}
where $\zeta, \eta, \omega, \chi$ are functions of null coordinates $u, v$ to be determined by the Einstein Maxwell equations.  $(\partial / \partial x)^a$ and $(\partial / \partial y)^a$ are spatial Killing vectors that describe the translation invariance of spacetime in the $x,y$ directions. 

A Newman-Penrose tetrad \cite{chandrasekhar1998mathematical} pertained to the spacetime metric in \eqref{eq:metric} may be defined as\cite{kuangEinsteinMaxwellSpacetimeTwo1999}
\begin{eqnarray}\label{eq:tetrad}
    l^a &= e^{-\zeta} \frac{\partial}{\partial u}, \\
    n^a &= e^{-\zeta} \frac{\partial}{\partial v}, \\
    m^a &= \frac{1}{\sqrt{2}} e^{-\eta} \sqrt{\cosh{\omega}} \left( e^{i\phi} \frac{\partial}{\partial x} - ie^{-\chi} \frac{\partial}{\partial y} \right), \\
    \Bar{m}^a &= \frac{1}{\sqrt{2}} e^{-\eta} \sqrt{\cosh{\omega}} \left( e^{-i\phi} \frac{\partial}{\partial x} + ie^{-\chi} \frac{\partial}{\partial y} \right).
\end{eqnarray}
where $\sin{\phi} = \tanh{\omega}$. Some relevant spin coefficients of the NP tetrad may be worked out to be \cite{kuangEinsteinMaxwellSpacetimeTwo1999}
\begin{equation}\begin{split}\label{eq:coefficients}
    \kappa &= \kappa' = \tau = \tau' = \alpha = \alpha' = 0, \\
    \rho &= -e^{-\zeta} \left( \frac{\partial \eta}{\partial u} + \frac{1}{2} \frac{\partial \chi}{\partial u} \right), \\
    \rho' &= -e^{-\zeta} \left( \frac{\partial \eta}{\partial v} + \frac{1}{2} \frac{\partial \chi}{\partial v} \right), \\
    \sigma &= \frac{1}{2} e^{-\zeta} \left( 1 + i\sinh{\omega} \right) \frac{\partial \chi}{\partial u} + \frac{i}{2} e^{-\zeta} \left( \frac{1}{\cosh{\omega}} + i\tanh{\omega} \right) \frac{\partial \omega}{\partial u}, \\
    \sigma' &= \frac{1}{2} e^{-\zeta} \left( 1 + i\sinh{\omega} \right) \frac{\partial \chi}{\partial v} + \frac{i}{2} e^{-\zeta} \left( \frac{1}{\cosh{\omega}} + i\tanh{\omega} \right) \frac{\partial \omega}{\partial v}.
\end{split}\end{equation}

Provided  the null congruence defined by ${\partial}/{\partial v}$ is shearfree so that $\sigma'=0$, the NP structure equations may be substantially  simplified and the metric coefficients in (\ref{metric}) are specified by a pair of functions 
$(f(u), g(v))$.
\begin{eqnarray}\label{g}
    g(v) = -|c|^2 v^2.
\end{eqnarray}
and $c$ is an arbitrary complex constant. 
 $f(u)$ is governed by a Sturm-Liouville type ODE given by 
\begin{eqnarray}\label{eq:f}
    \frac{\mathrm{d}^2 f}{\mathrm{d}u^2} + 2|\tilde{\sigma}|^2 f = 0,
\end{eqnarray}
where
\begin{eqnarray}
    \tilde{\sigma} = e^\zeta \sigma = \frac{1}{2} (1 + i\sinh{\omega}) \frac{\partial \chi}{\partial u} + \frac{1}{2} (\frac{1}{\cosh{\omega}} + i\tanh{\omega}) \frac{\partial \omega}{\partial u}.
\end{eqnarray}
The ordered pair $(f(u),\tilde\sigma)$ together with (\ref{g}) then generate an exact solution of the Einstein Maxwell equation. The parameters $(\eta, \chi, \zeta)$ are further subject to:
\begin{eqnarray}\label{chi}
    2\eta + \chi = \ln{(f(u) + g(v))},
\end{eqnarray}
and
\begin{eqnarray}\label{zeta}
    \zeta = -\frac{1}{4} \ln{(f(u) + g(v))}.
\end{eqnarray}

The NP equations only require $\omega$ and $\chi$ to be a function of $u$\cite{kuangEinsteinMaxwellSpacetimeTwo1999} and may be regarded as a free parameter in the spacetime metric. $\omega\ne 0$ means that the two Killing vectors  are not orthogonal to each other and this generates non-trivial polarisation for the gravitational wave (see \eqref{eq:Psi} below). $\chi$ describes the freedom in specifying the norm of  $(\partial / \partial y)^a$ and is governed by (\ref{chi}). 

\noindent  The spacetime metric in (\ref{metric}) takes the form 
 \begin{eqnarray}\label{eq:metric}
      \mathrm{d}s^2 &=& \frac{2}{(f - |c|^2 v^2)} \mathrm{d}u \mathrm{d}v - (f - |c|^2 v^2) \cosh{w} (e^{- \chi} \mathrm{d}x^2 + e^\chi \mathrm{d}y^2) \nonumber\\
   && - 2(f - |c|^2 v^2) \sinh{w} \mathrm{d}x \mathrm{d}y.
\end{eqnarray}
The non-zero Weyl curvature components are given by 
\begin{eqnarray}
    \Psi_0 &=& \frac{1}{(f - |c|^2 v^2)^{1/2}} \left( 2\frac{\mathrm{d} f}{\mathrm{d} u} + i(f - |c|^2 v^2) \left( \sinh{w} \frac{\partial \chi}{\partial u} - \frac{1}{\cosh{w}} \frac{\partial w}{\partial u} \right) \right) \tilde{\sigma} \nonumber\\
    &&+ (f - |c|^2 v^2)^{1/2} \frac{\partial \tilde{\sigma}}{\partial u} \label{eq:Psi}, \\
    \Psi_2 &=& \frac{|c|^2 v}{2 (f - |c|^2 v^2)^{3/2}} \frac{\mathrm{d} f}{\mathrm{d} u}.
\end{eqnarray}
The NP components of the Maxwell field are given as:
\begin{eqnarray}
    \phi_0 = \frac{c v}{(f - |c|^2 v^2)^{1/4}} \tilde{\sigma} e^{-i F},\quad \phi_2 = \frac{c}{(f - |c|^2 v^2)^{1/4}} e^{-i F}, \label{eq:phi}
\end{eqnarray}
where 
\begin{eqnarray}
    F = \int \left( \sinh{\omega} \frac{\partial \chi}{\partial u} + \frac{1}{\cosh{\omega}} \frac{\partial \omega}{\partial u} \right) \mathrm{d}u.
\end{eqnarray}

\section{Theoretical analysis}
\label{sec:resonance}

In this section we will undertake further analysis  of the structure of spacetime whose metric is given in (\ref{metric}). 
This will provide a basis on which we build up the numerical waveform in the next section. In what follows, we shall consider the simple case in which $\omega=0$.

\subsection{Parametric Resonance and Liapounov Instability}

Among the solutions which satisfy \eqref{eq:f}, there is a class of solutions involving the scattering of gravitational waves and electromagnetic waves which is of particular physical interest to us.  Consider 
\begin{eqnarray}\label{equ:sigma}
    2|\tilde{\sigma}|^2 =\sigma_0^2 (1 + h \cos{\gamma u})
\end{eqnarray}
which is 
a simple harmonic potential of frequency $\sigma_0$ subject to a sufficently small periodic perturbation. Here $0\le h \ll 1$ is a constant,  $\gamma = 2\sigma_0 + \varepsilon$ (see for instance \cite{landauMechanics2001}, chapter V section 27) where $\varepsilon$ is a sufficiently small detuned parameter so that the periodic perturbation becomes a  sideband deviated slightly from $2\sigma_0$.

Written as a first order system of ODE, \eqref{eq:f} may be given as 
\begin{equation}\begin{split}
       \frac{\mathrm{d}}{\mathrm{d} u} 
    \begin{pmatrix}
        f(u) \\ \xi(u)
    \end{pmatrix} = \mathbf{A} 
    \begin{pmatrix}
        f(u) \\ \xi(u)
    \end{pmatrix}
   \end{split}\end{equation}
where $\xi = \mathrm{d} f / \mathrm{d} u$ and
$$
    \mathbf{A} =
    \begin{pmatrix}
        0 & 1 \\
        -\tilde{\sigma}^2 & 0
    \end{pmatrix}.
$$
Provided $|\tilde{\sigma}| > 1$, 
 $|A^n| = \tilde{\sigma}^{2n} \rightarrow \infty$ when $n \rightarrow \infty$, which means that $f(u)$ is divergent as $ u\rightarrow \infty$.\cite{arnoldMathematicalMethodsClassical1997}. 
 It is well known that this is a simple case of  Liapounov instability  when (\ref{eq:f}) is looked on from a dynamical system perspective.

The aim of the present work is to work out the implications of this Liapounov instability on the propagation of gravitational wave characterised by the Weyl curvature component
$\Psi_0$. To this end, define $q = |\sigma_0|^2 h / 16$,  \eqref{eq:f} and together with \eqref{equ:sigma}  may be written as the standard Mathieu equation\cite{whittakerCourseModernAnalysis1996}:
\begin{eqnarray}\label{equ:Mathieu}
    \frac{\mathrm{d}^2 f}{\mathrm{d} u^2} + (|\sigma_0|^2 + 16q \cos{\gamma u}) f = 0.
\end{eqnarray}
A periodic solution with period $ 2\pi / \gamma$ may be given as 
\begin{eqnarray}\label{equ:f basis}
    f(u) = ae^{su} ce_1(u,q) + be^{su} se_1(u,q). 
\end{eqnarray}
$ce_1(u,q),se_1(u,q)$ are two periodic functions known as Mathieu function given as
\begin{equation}\begin{split}\label{equ:Mathieu basis}
    ce_1 (u,q) = \cos{(\frac{\gamma u}{2})} + q \cos{(\frac{3\gamma u}{2})} + \mathcal{O}(q^2), \\
    se_1 (u,q) = \sin{(\frac{\gamma u}{2})} + q \sin{(\frac{3\gamma u}{2})} + \mathcal{O}(q^2).
\end{split}\end{equation}
where $s$ is a positive constant defined by 
\begin{eqnarray}
    s^2 = \frac{1}{4} [(\frac{1}{2}h \sigma_0)^2 - \varepsilon^2].
\end{eqnarray}
We further assume $-\sigma_0 h / 2 < \varepsilon < \sigma_0 h / 2$ so that the system described in \eqref{equ:f basis} is Liapounov unstable.

\begin{figure}[htbp]
    \centering
    \includegraphics[width=0.55\textwidth]{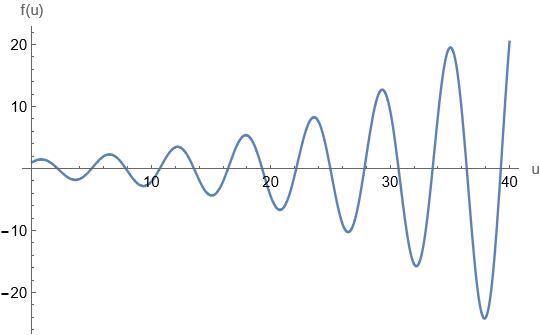}
    \caption{The graph of $f(u)$ at $\varepsilon = 0.2,\ \sigma_0 = 1,\ h = 0.5$. It is  a periodic function with the exponential function as envelope.}
    \label{Fig:Function} 
\end{figure}

In what follows, we shall work only with the zero order term in the perturbative expansion in \eqref{equ:Mathieu basis}, with terms involving the small parameter $q$ left out. Figure.\ref{Fig:Function} displays the variation of  $f(u)$ with respect to u.
We shall seek to justify numerically this step  towards the end of  next section.

\subsection{Scattering of gravitational waves with electromagnetic waves}

Given $(f,g)$ defined in \eqref{equ:f basis} and \eqref{g}, we shall seek to construct a spacetime that describes the scattering of gravitational and electromagnetic waves. 
To this end, further  refine the definition of $(f,g)$ as
\begin{equation}
    g(v) = -|c|^2v^2 \theta(v)
\end{equation}
where $\theta(v)$ is the step function that can be regarded as influencing $|c|$ by making it such that:
\begin{equation}\nonumber
    |c| = \left\{
    \begin{aligned}
        &|c|,\ \text{for} \ v > 0 \\
        &0,\ \text{for} \ v \leqslant 0
    \end{aligned}
    \right.
\end{equation}
and
\begin{equation}
    \tilde{f}(u) = \left\{
    \begin{aligned}
        &f(0)\ \text{for} \ u < 0 \\
        &f(u),\ \text{for} \ u \geqslant 0
    \end{aligned}
    \right.
\end{equation}

\begin{figure}[H] 
    \centering
    \includegraphics[width=0.55\textwidth]{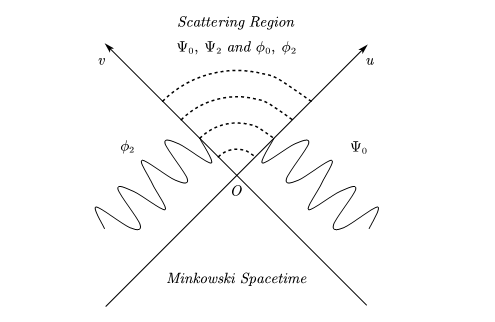} 
    \caption{Scattering picture of gravitational and electromagnetic waves. In region $u \leqslant 0, v \leqslant 0$, spacetime is Minkowskian; in region $u \leqslant 0, v \geqslant 0$, spacetime contains electromagnetic field with only $\phi_2$ component; $u \geqslant 0, v\leqslant 0$ , spacetime only contains gravitational wave. And gravitational wave and full coupling between gravitational and electromagnetic waves takes place at $u, v \geqslant 0$.}
    \label{Fig:Interaction} 
\end{figure}

Putting these into the metric in \eqref{eq:metric}, we then have:
\begin{enumerate}
    \item For $u \leqslant 0, v \leqslant 0$, spacetime is Minkowskian after proper normalization with parameter $f(0)$ choosing;
    \item $u \leqslant 0, v \geqslant 0$ is a spacetime generated by a propagating electromagnetic field with only $\phi_2$ component and vanishing Weyl curvature; 
    \item $u \geqslant 0, v \leqslant 0$ is a spacetime with only gravitational wave; 
    \item Full coupling between gravitational and electromagnetic field takes place in the region $u \geqslant 0, v\geqslant 0$ (see figure.\ref{Fig:Interaction}). 
\end{enumerate}

\subsection{Handling of spacetime singularity.}

From \eqref{eq:Psi}, it may be seen that there is a curvature singularity whenever $f - |c|^2 v^2=0$. The periodic nature of $f(u)$ (see \eqref{equ:f basis}) means that singularities will also appear
periodically at constant $v$. Further investigation reveals that this is not the spacetime singularity considered in the Hawking-Penrose singularity theorem \cite{hawking1970singularities,hawkingLargeScaleStructure1989}. The affine parameter $u$ of the null geodesic congruence defined by $\partial /\partial u$
is extendible beyond the singularity, in contrast to the existence of incomplete null geodesics ending abruptly considered in the Hawking-Penrose singularity theorem. 
The occurrence of curvature 
singularity in the present context originates from the divergent behaviour of null geodesics, as may be seen from the divergent behaviour of the spin coefficients
$\rho,\rho'$ in \eqref{eq:coefficients}. To avoid further complication due to the ill-behaviour of null geodesics and without compromising the physics we are interested in, we shall introduce a cutoff function $\theta(u) = \theta_1(u) + \theta_2(u)$ where
\begin{equation}
    \theta_1(u) = \left\{
    \begin{aligned}
        &1,\ u < u_i - \delta, \\
        &0,\ u \geqslant u_i - \delta,
    \end{aligned}
    \right.
    \ \text{and}\ 
    \theta_2(u) = \left\{
    \begin{aligned}
        &0,\ u < u_i + \delta, \\
        &1,\ u \geqslant u_i + \delta,
    \end{aligned}
    \right.
\end{equation}
where  $\delta$ is a sufficiently small real number, $u_i, i=1, \dots, n...$ are the  solutions of $f(u) = g(v)$ and $n$ is a natural number.  The gravitational waveform to be simulated numerically 
is essentially given by  $\theta(u) |\Psi_0(u)|$.  

The introduction of cutoff function means that the propagation of  $\Psi_0$ resembles a pulsed wave in electromagnetism (see figure \ref{Fig:cutoff} below). 
In electromagnetism, such an operation corresponds to a temporal filter to alter or attenuate the shape of a waveform. It is not clear how to
implement this physically in the context of gravitational wave physics. We will take it only as an permissible mathematical operation at present.

\section{Numerical simulation of gravitational waveform.}
\label{sec:sim}

In general, $\Psi_0$ is complex  and in what follows we shall numerically plot the modulus of these complex quantities and see how they propagate in the outgoing null direction described by the affine parameter $u$. This will enable us to see how intense beam of gravitational waves characterised by $\Psi_0$ is generated. If we take the modulus of $\Psi_0$ and $\phi_0, \phi_2$ as a measure of  gravitational and electromagnetic energy respectively, then we may see that an inter-exchange of energy between gravitational and electromagnetic energy takes place.

Consider $f(u)$ as defined in \eqref{eq:f} and set:
\begin{eqnarray}
    a = b = 1,\ \sigma_0 = 1, c = 2,\  g(v) = -4 v^2.
\end{eqnarray}
and $\varepsilon = 0.2$, $h = 0.5$ as perturbation.

\begin{figure}[t] 
    \centering
    \begin{subfigure}[t]{0.45\textwidth}
        \includegraphics[width=\textwidth]{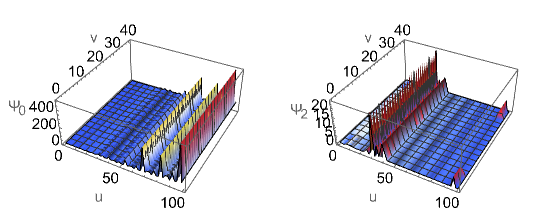}
        \subcaption{Propagation of $|\Psi_0|,\ |\Psi_2|$ in the $(u,v)$ plane. Along $u$ direction, the amplitude $|\Psi_0|$ grows exponentially.}
        \label{Fig:3D Psi}
    \end{subfigure}
    \hfill
    \begin{subfigure}[t]{0.45\textwidth}
        \includegraphics[width=\textwidth]{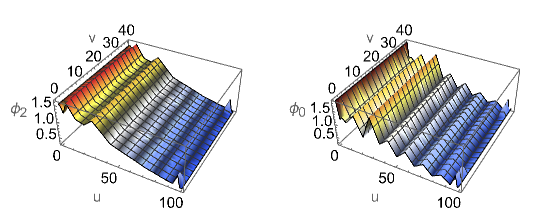}
        \subcaption{Propagation of $|\phi_0|,\ |\phi_2|$ in the $(u,v)$ plane. Compared with figure.\ref{Fig:3D Psi}, the value of $|\phi_2|$ decreases in tandem along $u$ direction with the same frequency.}
        \label{Fig:3D phi}
    \end{subfigure}
    \caption{Propagation of $\Psi_0,\Psi_2,\phi_0,\phi_2$ in the $(u,v)$ plane.}
    \label{Fig:3D}
\end{figure}

Figure.\ref{Fig:3D} describes the global variation of 
$|\Psi_0|, |\Psi_2|, |\phi_2|, |\phi_0|$ in the $(u,v)$ plane. As may be seen from the figure, the sharp increase in $|\Psi_0|$ is accompanied by the corresponding attenuation in $|\Psi_2|, |\phi_2|, |\phi_0|$. To understand better the increase of $|\Psi_0|$ in relation to $|\Psi_2|, |\phi_2|, |\phi_0|$, we will further take snapshot of $|\Psi_0|$  at constant $v$. 

\begin{figure}[t]
    \centering
    \begin{subfigure}[t]{0.28\textwidth}
        \includegraphics[width=\textwidth]{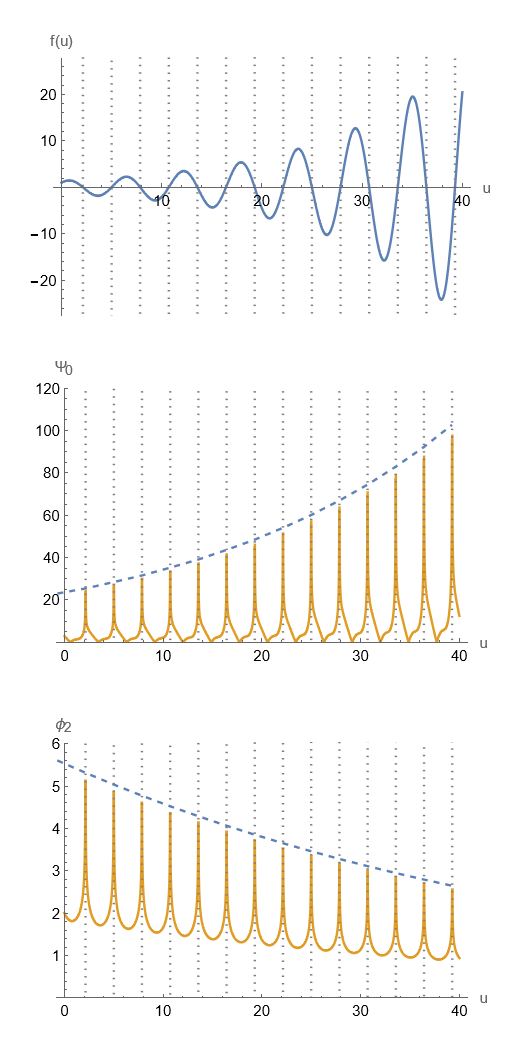} 
        \caption{Waveform at $v = 0.05$ slice, where $g(v) = -0.01$. Such slice is very close to $v = 0$ boundary slice.}
        \label{Fig:0-40 0.05}
    \end{subfigure}
    \hfill
    \begin{subfigure}[t]{0.28\textwidth}
        \includegraphics[width=\textwidth]{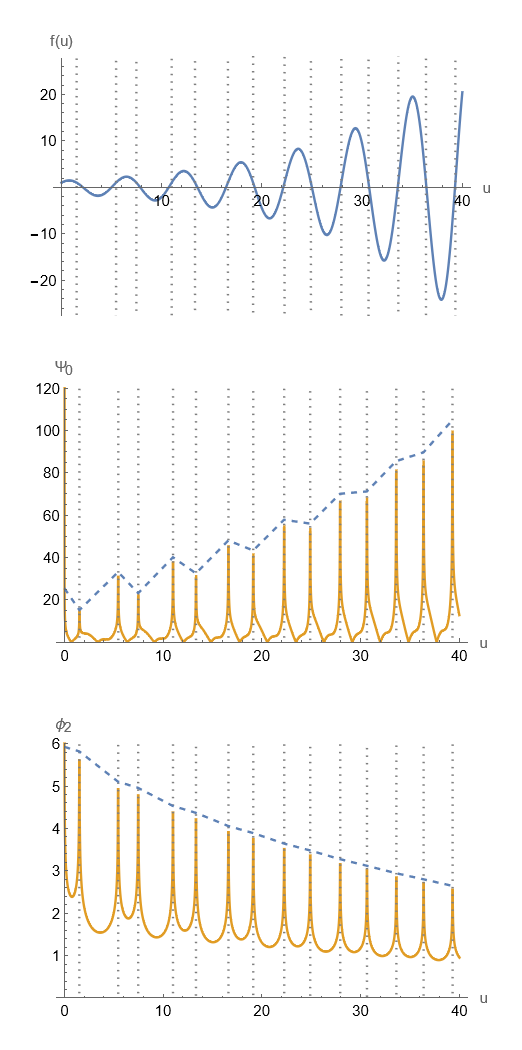} 
        \caption{Waveform at $v = 0.5$ slice, where $g(v) = -1$.}
        \label{Fig:0-40 0.5} 
    \end{subfigure}
    \hfill
    \begin{subfigure}[t]{0.28\textwidth}
        \includegraphics[width=\textwidth]{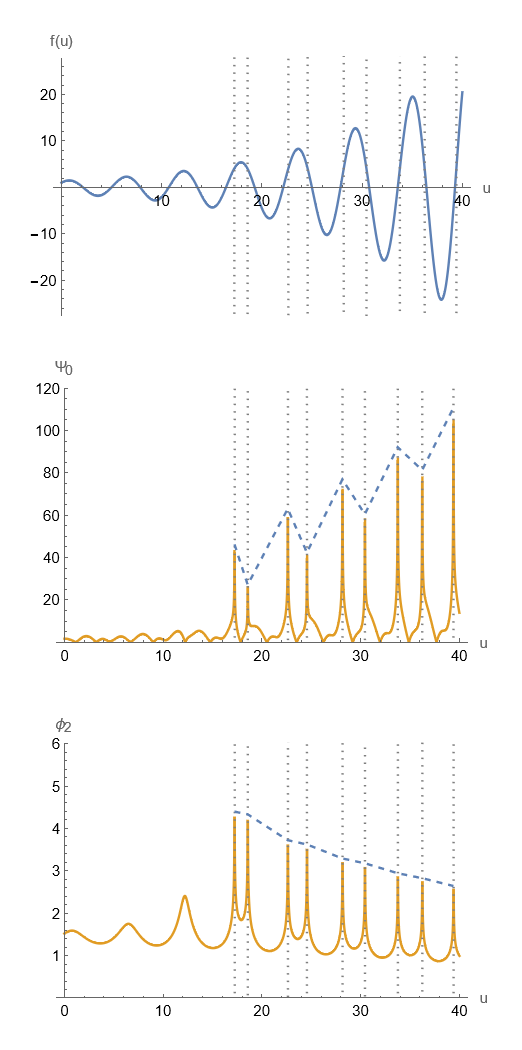} 
        \caption{Waveform at $v = 1$ slice, where $g(v) = -4$.}
        \label{Fig:0-40 1}
    \end{subfigure}
    \caption{Waveform of $|\Psi_0|$ at $\varepsilon = 0.2,\ \sigma_0 = 1,\ h = 0.5$ in the interval $u \in (0,40)$ with singularities cutoff in different $v = \hbox{constant}$ slice.  The vertical dotted lines represent the solutions of $f(u) = g(v)$, where the spacetime singularities are located. The dashed lines represent the envelope. Correlations may be observed between $\Psi_0$ and $\phi_2$. The larger $g(v)$ is, the later parametric resonance appears. However, the shape of the waveforms remain unchanged. }
    \label{Fig:0-40}
\end{figure}

From figure.\ref{Fig:0-40}, it may be seen that the exponential increase of the amplitude of $\Psi_0$ is at the expense of the corresponding decrease in amplitudes of $\phi_2$. $\Psi_0$ draws on energy not just from the electromagnetic field but also from the $\Psi_2$. When certain resonance condition is met, the pumping of electromagnetic energy is transferred to the large amplitude of $\Psi_0$. Further, the energy transfer is coherent in the sense that the increase of $\Psi_0$ and the decrease of $\phi_2$ are at the same frequency. Like the free electron laser, the frequency of  $\Psi_0$ is tunable by adjusting the periodicity of $\phi_0$ and $\phi_2$. In the vacuum limit when the Maxwell field is absent, the parametric resonance phenomenon persists, however this corresponds only  to the exchange of energy between the Weyl curvature components $\Psi_0$ and $\Psi_2$ of the gravity field. 

As the numerical resolution in figure.\ref{Fig:0-40} is not good enough to see the pulse nature of $\Psi_0$ due to the cutoff of singularities. In figure.\ref{Fig:cutoff} below, single period of the wave propagation is displayed to illustrated the pulsed nature of the gravitational wave propagation. 

\begin{figure}[H]
    \centering
    \includegraphics[width=0.4\textwidth]{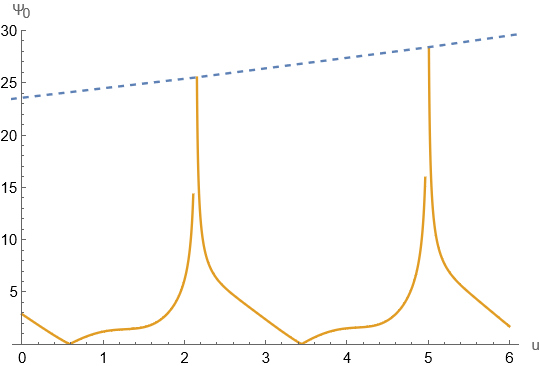} 
    \caption{Pulsed nature of gravitational wave propagation. }
    \label{Fig:cutoff} 
\end{figure}

\begin{figure}[H]
    \centering
    \includegraphics[width=0.4\textwidth]{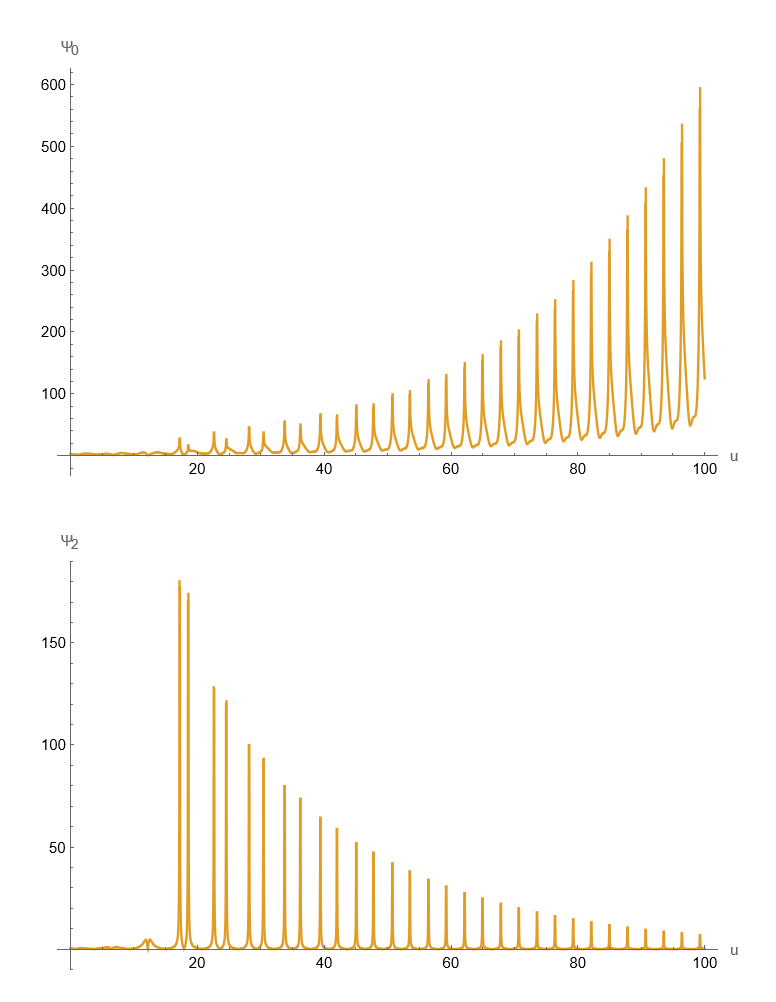}
    \caption{Propagation of $\Psi_0$ and $\Psi_2$ at $v = 1$ slice where $g(v) = -4$ is a delay function which determines the onset of instability at constant $v$. }
    \label{Fig:Psi_0 and Psi_2}
\end{figure}

\begin{figure}[H] 
    \centering
    \includegraphics[width=0.4\textwidth]{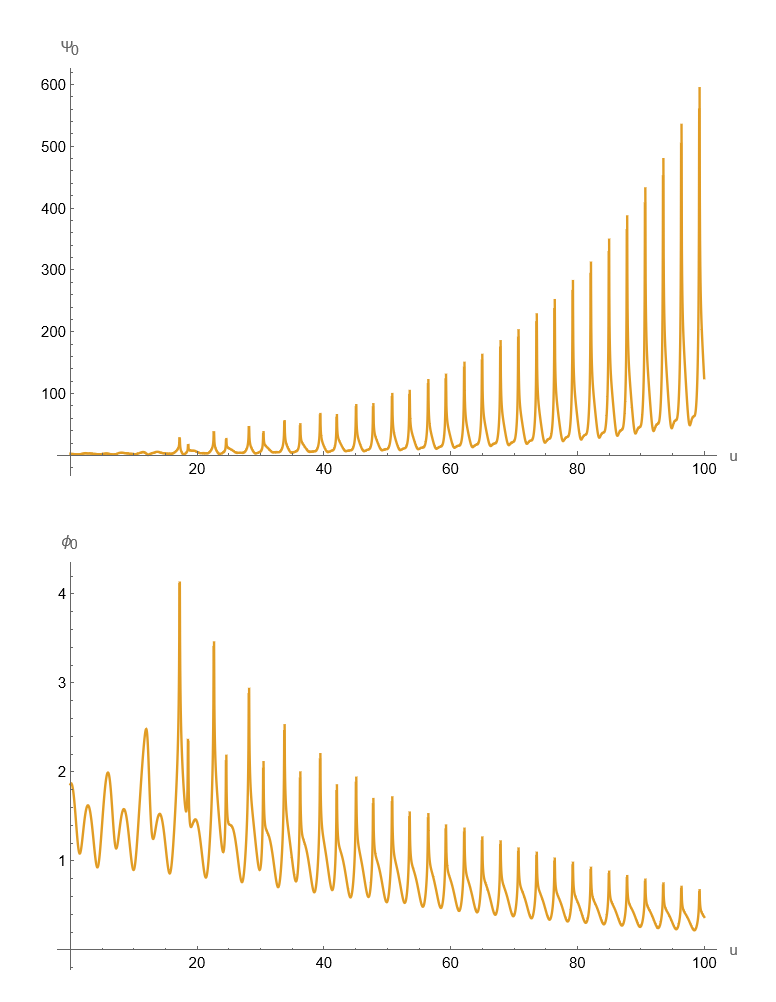}
    \caption{Propagation of $\Psi_0$ and $\phi_0$ at $v = 1$ slice where $g(v) = -4$.}
    \label{Fig:Psi_0 and phi_0}
\end{figure}

Next, we plot at $v = 1$ slice the exponential increase in $|\Psi_0|$ in relation to $\Psi_2, \phi_0$ in figure.\ref{Fig:Psi_0 and Psi_2} and figure.\ref{Fig:Psi_0 and phi_0} along the outgoing null direction $u$. From figure 4c (which displays the exponential increase of $|\Psi_0|$ in relation to $\phi_2$),figure.\ref{Fig:Psi_0 and Psi_2} and figure.\ref{Fig:Psi_0 and phi_0}, we may see that the function $g(v)$ determines the time $u$ at which instability is switched on. 

So far we have considered only an approximate solution to the Mathieu equation with the small parameter $q$ totally left out. To understand the possible contribution from terms
 involving  $q$ in \eqref{equ:Mathieu basis}, we repeat the numerical wavefrom simulations with terms up to second order in $q$ included in \eqref{equ:Mathieu basis}. No noticable difference in the waveform from that displayed in figure.\ref{Fig:0-40} is observed. The exponential 
amplification of gravitational wave amplitude is dominated by the zero order term given in \eqref{equ:Mathieu}. 

\section{Concluding Remarks}

The present work serves to point out, at least at the theoretical level, that an intense pulse gravitational wave train may be generated by the conversion of electromagnetic energy into gravitational energy, provided certain resonance condition is met, in some way similar to that for a free electron laser.  The class of examples considered is admittedly highly idealistic and more work remains to be done to explore this feasibility at a more practical level. It offers hope, however remote it may seem for the time being, that in future an intense beam of gravitational wave akin to a laser in electromagnetic theory may be generated  in a laboratory,  unlike the current situation when we only count on astrophysical sources for
gravitational wave detection. As far as detection of gravitational waves from an astrophysical source is concerned, it is conceivable that, provided we have in advance knowledge of the frequency of the wave to be detected, electromagnetic energy may be pumped in to enhance the amplitude of the gravitational waves so that the signal to noise ratio may be substantially enhanced \cite{braginskyGravitationalelectromagneticResonance1972}. This remains a feasibility to be explored in future.

\section*{Acknowledgements}

 The work is supported by the National Key Research and Development Program of China under Grant 2021YFC2202501. 

\section*{References}

\bibliographystyle{unsrt}
\bibliography{refs}

\end{document}